\documentclass[aps,prd,superscriptaddress,showpacs]{revtex4}
\usepackage{graphicx}
\usepackage{epstopdf}
\usepackage{amsmath}
\usepackage{amsfonts}
\usepackage{amssymb}
\usepackage{latexsym}

\begin{document}
\title{Remarks on the static potential driven by vacuum nonlinearities in $D=3$ models}
\author{Patricio Gaete} \email{patricio.gaete@usm.cl} 
\affiliation{Departamento de F\'{i}sica and Centro Cient\'{i}fico-Tecnol\'ogico de Valpara\'{i}so, Universidad T\'{e}cnica Federico Santa Mar\'{i}a, Valpara\'{i}so, Chile}
\author{Jos\'{e} A. Helay\"{e}l-Neto}\email{helayel@cbpf.br}
\affiliation{Centro Brasileiro de Pesquisas F\'{i}sicas (CBPF), Rio de Janeiro, RJ, Brasil} 
\date{\today}

\begin{abstract}
Within the framework of the gauge-invariant, but path-dependent, variables formalism, we study the manifestations of vacuum electromagnetic nonlinearities in $D=3$ models. For this we consider both  generalized Born-Infeld and Pagels-Tomboulis-like electrodynamics, as well as, an Euler-Heisenberg-like electrodynamics. We explicitly show that generalized Born-Infeld and Pagels-Tomboulis-like electrodynamics are equivalent, where the static potential profile contains a long-range (${\raise0.5ex\hbox{$\scriptstyle 1$}\kern-0.1em/\kern-0.15em\lower0.25ex\hbox{$\scriptstyle {{r^2}}$}}$-type) correction to the Coulomb potential. Interestingly enough, for an Euler-Heisenberg-like electrodynamics the interaction energy contains a linear potential, leading to the confinement of static charges. 
\end{abstract}
\pacs{14.70.-e, 12.60.Cn, 13.40.Gp}
\maketitle

\section{Introduction}

Physical consequences of vacuum electromagnetic nonlinearities such as vacuum birefringence and
vacuum dichroism have been a topic of increasing interest in recent times  \cite{Bamber,Burke,nphoton,Tommasini1,Tommasini2}. For example, the PVLAS experiment is one of the most sensitive test to search vacuum magnetic birefringence in the presence of an external magnetic field. However, despite remarkable progress, this has not yet been observed but it is hoped to reach this goal in the next few years \cite{DellaValle}. The advent of new laser facilities also has generated interest about this topic
by measuring the scattering of intense laser pulses \cite{Sarazin}. It is worth recalling, at this stage, that electromagnetic nonlinearity in vacuum is a quantum effect predicted by the Euler-Heisenberg effective Lagrangian for slowly varying electromagnetic fields \cite{Euler}. We also recall in passing that, in classical electrodynamics, vacuum electromagnetic properties are described by the vacuum permittivity (${\varepsilon _0}$)  and the vacuum permeability (${\mu _0}$).

Meanwhile, in previous works \cite{Nonlinear,Logarithmic,Nonlinear2} we have considered the physical consequences presented by different models of $(3+1)$-D nonlinear Electrodynamics in vacuum. In fact, we have shown that for Generalized Born-Infeld, and Logarithmic Electrodynamics the field energy of a point-like charge is finite. Moreover, Generalized Born-Infeld, Exponential, Logarithmic and Massive Euler-Heisenberg-like Electrodynamics display the vacuum birefringence phenomenon. The point we wish to emphasize, however, is that all these Electrodynamics share a long-range correction (${\raise0.5ex\hbox{$\scriptstyle 1$}
\kern-0.1em/\kern-0.15em\lower0.25ex\hbox{$\scriptstyle {{r^5}}$}}$-type) to the Coulomb potential.

On the other hand, as is well known, a full understanding of the QCD vacuum structure and color confinement mechanism remains still elusive. However, phenomenological models still represent an interesting framework for understanding features of the confinement physics \cite{Nambu,'t Hooft, Mandelstam}. Incidentally, it is of interest to recall here the illustrative scenario of dual superconductivity, where it is conjectured that the QCD vacuum behaves as a dual-type II superconductor. Accordingly, because of the condensation of magnetic monopoles, the the color electric flux linking quarks is squeezed into strings, and the non-vanishing string tension represents the proportionality constant in the linear potential. In this context, we also recall the Pagels-Tomboulis model \cite{Pagels}, which was introduced to reproduce the trace anomaly from QCD. Interestingly, when this model is considered with a dilaton field, the interaction energy contains a linear confining potential  \cite{Wereszcz}. Another relevant model is the one-loop effective action for QCD in $(2+1)$-dimensions, where this new QCD vacuum acts like a dielectric medium and leads to confinement \cite{Dittrich}. 

With these considerations in mind, the purpose of this work is to further elaborate on the physical content of electromagnetic nonlinearities on a physical observable. To do this, we shall work out the static potential for the different three-dimensional field theoretic models along the lines of \cite{Nonlinear,Logarithmic,Nonlinear2}. The advantage of using this development lies in the fact that the interaction energy between two static charges is obtained once a judicious identification of the physical degrees of freedom is made \cite{Pato1,Pato2}. As will be seen, Born-Infeld-like electrodynamics and Pagels-Tomboulis electrodynamics are equivalent. While a three-dimensional Euler-Heisenberg-like electrodynamics is analogous to that encountered in both a three-dimensional gluodynamics and in a three-dimensional Yang-Mills with spontaneous symmetry breaking of scale symmetry in the short distance regime \cite{PatJose}. In other terms, in this work we are concerned with the physical content associated with duality, where duality refers to a an equivalence between two or more quantum field theories whose corresponding classical theories are different. In this sense it should be understood our equivalence among theories.
 
Our work is organized according to the following outline: in Section II, we consider Born-Infeld-like electrodynamics and compute the interaction energy for a fermion-antifermion pair, the calculation shows a long-correction (${\raise0.5ex\hbox{$\scriptstyle 1$}
\kern-0.1em/\kern-0.15em\lower0.25ex\hbox{$\scriptstyle {{r^2}}$}}$-type) to the Coulomb potential. 
In Section III, we extend our analysis for the Pagels-Tomboulis electrodynamics and for a three-dimensional Euler-Heisenberg-like electrodynamics. Interestingly enough, for Euler-Heisenberg-like electrodynamics, the static potential profile contains a linear term, leading to the confinement of static charges. Finally, some concluding remarks are made in Sec. IV. 

In our conventions the signature of the metric is ($+1,-1,-1$).

\section{On Born-Infeld-like electrodynamics} 

As already mentioned, the main focus of our work is to calculate explicitly the interaction energy between static point-like sources for three electrodynamics. With this purpose, let us consider first a Born-Infeld-like electrodynamic in $(2+1)$ dimensions. This would not only provide the theoretical setup for our subsequent work, but also fix the notation. The initial point of our analysis is the Lagrangian density:   
\begin{equation}
{\cal L} = \beta ^2 \left\{ {1 - \left[ {1 + \frac{2}{{\beta ^2 }}{\cal F} - \frac{1}{{\beta ^4}}{\cal G}^2 } \right]^p } \right\}, \label{BIL05}
\end{equation}
where ${\cal F} = \frac{1}{4}F_{\mu \nu } F^{\mu \nu }$, $   
{\cal G} = \frac{1}{4}F_{\mu \nu } \tilde F^{\mu \nu }$, $F_{\mu \nu }  = \partial _\mu  A_\nu   - \partial _\nu  A_\mu$ and $\tilde F^{\mu \nu }  = \frac{1}{2}\varepsilon ^{\mu \nu \rho \lambda } F_{\rho \lambda }$ is the dual electromagnetic field strength tensor. 
As we have already explained in \cite{Nonlinear,Logarithmic,Nonlinear2}, we shall confine ourselves to the domain $0 < p < 1$.

In the case under consideration it follows that:
\begin{equation}
\partial _\mu  \left[ {\frac{1}{\Gamma^{1-p} }\left( {F^{\mu \nu }  - \frac{1}{{\beta ^2 }}{\cal G}\tilde F^{\mu \nu } } \right)} \right] = 0, \label{BIL10}
\end{equation}
while the Bianchi identity is given by
\begin{equation}
\partial _\mu  \tilde F^{\mu \nu }  = 0, \label{BIL15}
\end{equation}
where $\Gamma  = 1 + \frac{{{2\cal F}}}{{\beta ^2 }} - \frac{{{\cal G}^2 }}{{\beta ^4}}$. 

It should be further noted that Gauss' law reduces to,
\begin{equation}
\nabla  \cdot {\bf D} = 0, \label{BIL20}
\end{equation}
where $\bf D$ is given by
\begin{equation}
{\bf D} = \frac{{{\bf E} + \frac{1}{{\gamma ^2 }}\left( {{\bf E} \cdot {\bf B}} \right){\bf B}}}{{[1 - \frac{{\left( {{\bf E}^2  - {\bf B}^2 } \right)}}{{\beta ^2 }} - \frac{1}{{\beta ^4}}\left( {{\bf E} \cdot {\bf B}} \right)^2]^{1-p} }}. \label{BIL25}
\end{equation}
It can easily be seen that, for $J^0 (t,{\bf r}) = e\delta ^{\left( 2 \right)} \left( {\bf r} \right)$, the ${\bf D}$-field lies along the radial direction and is given by ${\bf D} = \frac{Q}{{r }}\hat r$, where $Q = \frac{e}{{2\pi }}$. It is also important to observe that for a point-like charge, e, at the origin, the expression
\begin{equation}
\frac{Q}{{r }} = \frac{{|{\bf E}|}}{{\left( {1 - \frac{{{\bf E}^2 }}{{\beta ^2 }}} \right)^{1 - p} }}, \label{BIL30}
\end{equation}
shows that, for $r \to 0$, the electrostatic field is regular at the origin (where it acquires its maximum, $ 
|{{\bf E}_{\max }}|= \beta$) only with $p < 1$. As we have noted before, the $p < 0$ case is excluded because there could exist field configurations for which the Lagrangian density would blow up. While for $p > 1$, $|{\bf E}|$ becomes singular at $r=0$. Thus, in this work we shall concentrate once again in the $0 < p < 1$ case.

With these considerations in mind, we shall now discuss the interaction energy between static point-like sources for the model under study. To this end, we will calculate the expectation value of the energy operator $ H$ in the physical state $ |\Phi\rangle$, along the lines of Refs. \cite{Nonlinear,Logarithmic,Nonlinear2}. The starting point is the Lagrangian density (\ref{BIL05}) with $p = {\raise0.5ex\hbox{$\scriptstyle 3$}\kern-0.1em/\kern-0.15em\lower0.25ex\hbox{$\scriptstyle 4$}}$, 
\begin{equation}
{\cal L} = \beta ^2 \left\{ {1 - \left[ {1 + \frac{1}{{3\beta ^2 }}F_{\mu \nu }^2  - \frac{1}{{24\beta ^4}}\left( {F_{\mu \nu } \tilde F^{\mu \nu } } \right)^2 } \right]^{{\raise0.5ex\hbox{$\scriptstyle 3$}
\kern-0.1em/\kern-0.15em
\lower0.25ex\hbox{$\scriptstyle 4$}}} } \right\},  \label{BIL35}
\end{equation}
where we have redefined the coefficients in front of ${\cal F}$ and ${\cal G}$.

As we have indicated in \cite{Nonlinear,Logarithmic,Nonlinear2}, to handle the exponent $  
{\raise0.5ex\hbox{$\scriptstyle 3$}
\kern-0.1em/\kern-0.15em
\lower0.25ex\hbox{$\scriptstyle 4$}}$ in expression (\ref{BIL35}), we incorporate an auxiliary field $v$ such that its equation of motion gives back the original theory. Therefore the corresponding Lagrangian density takes the form
\begin{equation}
{\cal L} = \beta ^2  - 3\beta ^2 v - vF_{\mu \nu }^2  + \frac{v}{{8\beta ^2 }}\left( {F_{\mu \nu } \tilde F^{\mu \nu } } \right)^2  - \frac{{\beta ^2 }}{{4^4 }}\frac{1}{{v^3 }}. \label{BIL40}
\end{equation}

It is worthwhile sketching at this point the canonical quantization of this theory from the Hamiltonian analysis point of view. It may now easily be verified that the canonical momenta are $    
\Pi ^\mu   =  - 4v\left( {F^{0\mu }  - \frac{1}{{4\beta ^2 }}F_{\alpha \beta } \tilde F^{\alpha \beta } \tilde F^{0\mu } } \right)$, so one immediately identifies the two primary constraints 
$\Pi ^0  = 0$ and $p \equiv \frac{{\partial L}}{{\partial \dot v}} = 0$. Furthermore, the momenta are ${\Pi _i} = 4v{D_{ij}}{E_j}$. Here ${E_i} = {F_{i0}}$ and  ${D_{ij}} = {\delta _{ij}} + \frac{1}{{{\beta ^2}}}{B_i}{B_j}$. From this it follows that electric field can be written as ${E_i} = \frac{1}{{4v\det D}}\left( {{\delta _{ij}}\det D - \frac{1}{{{\beta ^2}}}{B_i}{B_j}} \right){\Pi _j}$. In such a case, the canonical Hamiltonian reduces to
\begin{equation}
H_C  = \int {d^2 x} \left\{ {\Pi _i \partial ^i A^0  + \frac{1}{{8v}}{\bf \Pi} ^2  - \beta ^2  + 3\beta ^2 v + 2v{\bf B}^2  + \frac{{\beta ^2 }}{{4^4 }}\frac{1}{{v^3 }} - \frac{{\left( {{\bf \Pi}  \cdot {\bf B}} \right)^2 }}{{8v\beta ^2 \left( {1 + \frac{{{\bf B}^2 }}{{\gamma ^2 }}} \right)}}} \right\}. \label{BIL45}
\end{equation}
Next, we also notice that by requiring the primary constraint $\Pi^{0}$ to be preserved in time, one obtains the secondary constraint $\Gamma _1  = \partial _i \Pi ^i  = 0$. Similarly for the constraint $p$, we get the auxiliary field $v$ as 
\begin{eqnarray}
v = \frac{1}{{\sqrt {48{\beta ^4}\det D} }}\sqrt {{\beta ^2}{\Pi ^2}\det D + \sqrt {{{\left( {{\beta ^2}{\Pi ^2}\det D} \right)}^2} + 9{\beta ^8}{{\left( {\det D} \right)}^2}} }, \label{BIL50}
\end{eqnarray}
which will be used to eliminate $v$. We observe that to get this last expression we have ignored the magnetic field in equation (\ref{BIL45}), because it add nothing to the static potential calculation, as we will show it below. According to usual procedure, the corresponding total Hamiltonian that generates the time evolution of the dynamical variables is $  
H = H_C  + \int {d^2 x} \left( {u_0(x) \Pi_0(x)  + u_1(x) \Gamma _1(x) } \right)$, where $u_o(x)$ and $u_1(x)$ are the Lagrange multiplier utilized to implement the constraints. It is a simple matter to verify that $\dot A_0 \left( x \right) = \left[ {A_0 \left( x \right),H} \right] = u_0 \left( x \right)$, which is an arbitrary function. Since $\Pi^0=0$ always, neither $A^0$ nor $\Pi^0$ are of interest in describing the system and may be discarded from the theory. Hence, we can write
\begin{equation}
H= \int {d^2 x} \left\{ {w(x)\partial ^i \Pi _i + \frac{1}{{8v}}{\bf \Pi} ^2  - \beta ^2  + 3\beta ^2 v + 
\frac{{\beta ^2 }}{{4^4 }}\frac{1}{{v^3 }} }\right\}. \label{BIL55}
\end{equation}
where $w(x) = u_1 (x) - A_0 (x)$ and $v$ is given by (\ref{BIL50}).

We can at this stage impose a gauge condition, so that in conjunction with the constraint ${\Pi ^0} = 0$, it is rendered into a second class set. A particularly convenient choice is
\begin{equation}
\Gamma _2 \left( x \right) \equiv \int\limits_{C_{\xi x} } {dz^\nu
} A_\nu \left( z \right) \equiv \int\limits_0^1 {d\lambda x^i }
A_i \left( {\lambda x} \right) = 0. \label{BIL60}
\end{equation}
where  $\lambda$ $(0\leq \lambda\leq1)$ is the parameter describing
the spacelike straight path $ z^i = \xi ^i  + \lambda \left( {x -\xi } \right)^i $, and $ \xi $ is a fixed point (reference point). We also recall that there is no essential loss of generality if we restrict our considerations to $ \xi ^i=0 $. Hence the only nontrivial Dirac bracket for the canonical variables is given by
\begin{equation}
\left\{ {A_i \left( x \right),\Pi ^j \left( y \right)} \right\}^ *
= \delta _i^j \delta ^{\left( 2 \right)} \left( {x - y} \right) -
\partial _i^x \int\limits_0^1 {d\lambda x^i } \delta ^{\left( 2
\right)} \left( {\lambda x - y} \right). \label{BIL65}
\end{equation}

We now proceed to compute the interaction energy for the model under consideration. As mentioned above, to do that we need to compute the expectation value of the energy operator $H$ in the physical state $\left| \Phi  \right\rangle$. Following Dirac \cite{Dirac}, we write the physical state $\left| \Phi  \right\rangle$ as
\begin{equation}
\left| \Phi  \right\rangle  \equiv \left| {\bar \Psi ({\bf y})\Psi ({{\bf y}^ \prime })} \right\rangle  = \bar \psi ({\bf y})\exp (ie\int_{{{\bf y}^ \prime }}^{\bf y} {d{z^i}{A_i}(z)} )\psi ({{\bf y}^ \prime })\left| 0 \right\rangle,                              
\label{BIL70}
\end{equation}
where $\left| 0 \right\rangle$ is the physical vacuum state and the
line integral appearing in the above expression is along a spacelike
path starting at ${\bf y}\prime$ and ending at $\bf y$, on a fixed
time slice. The above expression clearly shows that, each of the states $(\left| \Phi  \right\rangle)$, 
represents a fermion-antifermion pair surrounded by a cloud of gauge fields to maintain gauge invariance.

Taking the above Hamiltonian structure into account, we see that
\begin{equation}
\Pi _i \left( x \right)\left| {\overline \Psi \left( \mathbf{y }\right)\Psi
\left( {\mathbf{y}^ \prime } \right)} \right\rangle = \overline \Psi \left( 
\mathbf{y }\right)\Psi \left( {\mathbf{y}^ \prime } \right)\Pi _i \left( x
\right)\left| 0 \right\rangle + e\int_ {\mathbf{y}}^{\mathbf{y}^ \prime } {\
dz_i \delta ^{\left( 2 \right)} \left( \mathbf{z - x} \right)} \left| \Phi
\right\rangle.  \label{BIL75}
\end{equation}
Therefore, the lowest-order modification in ${\beta ^2}$ of the interaction energy can be written as
\begin{equation}
{\left\langle H \right\rangle _\Phi } = {\left\langle H \right\rangle _0} + {V_1} + {V_2}, \label{BIL80}
\end{equation}
where $\left\langle H \right\rangle _0  = \left\langle 0 \right|H\left| 0 \right\rangle$. The $V_1$, $V_2$ terms are given by
\begin{equation}
{V_1} =  - \frac{1}{2}\left\langle \Phi  \right|\int {{d^2}x} {\Pi _i}{\Pi ^i}\left| \Phi  \right\rangle, \label{BIL85} 
\end{equation}
and
\begin{equation}
{V_2} =  - \frac{1}{{12{\beta ^2}}}\left\langle \Phi  \right|\int {{d^2}x} {\Pi ^4}\left| \Phi  \right\rangle. \label{BIL90} 
\end{equation}
Using equation (\ref{BIL75}) and following our earlier procedure, we see that the potential for two opposite charges, localized at ${\bf y}$ and ${\bf {y^\prime}}$, takes the form   
\begin{equation}
V = \frac{{{e^2}}}{{4\pi }}\log \left( {\frac{L}{{{L_0}}}} \right) + \frac{{{e^4}}}{{192{\beta ^2}{\pi ^2}}}\frac{1}{{{L^2}}},
\label{BIL95}
\end{equation}
where $|{\bf y} - {{\bf y}^ \prime }| = L$ and $L_0$ is a cut-off. It should be further noted that the present cut-off $L_0$ is putting by hand. We shall come back to this point below.

Before concluding this subsection it is constructive to briefly examine an alternative derivation of our previous result, which permits us to check the internal consistency of our procedure. In order to illustrate the discussion, we begin by recalling that
\begin{equation}
V \equiv \frac{e}{2}\left( {{\cal A}_0 \left( {\bf 0} \right) - {\cal A}_0 \left( {\bf L} \right)} \right), \label{BIL100}
\end{equation}
where the physical scalar potential is given by
\begin{equation}
{\cal A}_0 (t,{\bf r}) = \int_0^1 {d\lambda } r^i E_i (t,\lambda
{\bf r}). \label{BIL105}
\end{equation}
This equation follows from the vector gauge-invariant field expression
\begin{equation}
{\cal A}_\mu  (x) \equiv A_\mu  \left( x \right) + \partial _\mu  \left( { - \int_\xi ^x {dz^\mu  A_\mu  \left( z \right)} } \right), \label{BIL110}
\end{equation}
where the line integral is along a spacelike path from the point $\xi$ to $x$, on a fixed slice time. It should again be stressed here that the gauge-invariant variables (\ref{BIL110}) commute with the sole first constraint (Gauss law), showing in this way that these fields are physical variables. 

It should be noted that Gauss' law for the present theory reads
\begin{equation}
{\partial _i}{\Pi ^i} = {J^0},  \label{BIL111}
\end{equation}
where ${\Pi ^i} = 4v{E^i}$ and $v$ is given by equation (\ref{BIL50}). Note that we have included the external current $J^0$ to represent the presence of external charges. Since our interest here is in estimating the lowest-order correction to the Coulomb energy, we will retain only the leading quadratic term in expression  (\ref{BIL111}). In such a case, for ${J^0}({\bf r}) = e{\delta ^{\left( 2 \right)}}\left( {\bf r} \right)$, the electric field reduces to  
\begin{equation}
{\bf E} = \sqrt 3 \beta \frac{e}{{2\pi }}\frac{1}{{\sqrt {{{\left( {{\raise0.5ex\hbox{$\scriptstyle e$}
\kern-0.1em/\kern-0.15em
\lower0.25ex\hbox{$\scriptstyle {2\pi }$}}} \right)}^2} + \sqrt {{{\left( {{\raise0.5ex\hbox{$\scriptstyle e$}
\kern-0.1em/\kern-0.15em
\lower0.25ex\hbox{$\scriptstyle {2\pi }$}}} \right)}^4} + 9{\beta ^2}{r^4}} } }}\hat r. \label{BIL112}
\end{equation}

Using (\ref{BIL112}), we can express (\ref{BIL105}) as
\begin{equation}
{{\cal A}_0}(t,{\bf r}) =  - \frac{e}{{2\pi r}}\int_0^1 {d\lambda } \frac{1}{{\sqrt {{\lambda ^2} + {a^2}} }}, \label{BIL115}
\end{equation}
where ${a^2} = {\left( {\frac{e}{{2\pi \beta }}} \right)^2}\frac{1}{{3{r^2}}}$. We can, therefore, write
\begin{equation}
{{\cal A}_0}(t,{\bf r}) = - \frac{e}{{4\pi }}\log \left( {\frac{L}{{{L_0}}}} \right) - \frac{{{e^4}}}{{192{\pi ^3}{\beta ^2}}}\frac{1}{{{L^2}}}, \label{BIL120}
\end{equation}
where the cut-off $L_0$ is given by ${L_0} = \frac{e}{{2\sqrt 3 \pi \beta }}$. Notice that, in contrast to the previous calculation, the cut-off is completely determined.

Accordingly, by employing Eq. (\ref{BIL100}), finally we end up with the potential for a pair of static point-like opposite charges located at $\bf 0$ and $\bf L$, 
\begin{equation}
V = \frac{{{e^2}}}{{4\pi }}\log \left( {\frac{L}{{{L_0}}}} \right) + \frac{{{e^4}}}{{192{\pi ^3}{\beta ^2}}}\frac{1}{{{L^2}}}, \label{BIL125}
\end{equation}
after subtracting a self-energy term.

One immediately sees that that Born-Infeld-like electrodynamics in $(2+1)$-dimensions also has a rich structure reflected by its long-range correction to the Coulomb potential.

\section{On Euler-Heisenberg-like electrodynamics}

\subsection{Abelian Pagels-Tomboulis model}
We shall now discuss the interaction energy between static point-like sources for the Abelian form of the Pagels-Tomboulis-like model \cite{Pagels}. Proceeding in the same way as we did in the previous section we will compute the expectation value of the energy operator $H$ in the physical state $\left| \Phi  \right\rangle$. We commence our discussion by considering:
\begin{equation}
{\cal L} =  - \frac{1}{4}{\left( {-\frac{{{F_{\mu \nu }}{F^{\mu \nu }}}}{{2{\Lambda ^4}}}} \right)^{2\delta }}{F_{\mu \nu }}{F^{\mu \nu }}, \label{EHM05}
\end{equation}
where $\Lambda$ is a dimensional and $\delta$ is a dimensionless constant. Next, equation (\ref{EHM05}) can be written alternatively in the form 
\begin{equation}
{\cal L} =  - \frac{1}{4}{F_{\mu \nu }}{F^{\mu \nu }} - \frac{\delta }{2} {F_{\mu \nu }}{F^{\mu \nu }}\ln \left( -{\frac{{{F_{\mu \nu }}{F^{\mu \nu }}}}{{2{\Lambda ^4}}}} \right),  \label{EHM10}
\end{equation}
where to get the last line we used $2\delta \ln \left(- {\frac{{{F_{\mu \nu }}{F^{\mu \nu }}}}{{2{\Lambda ^4}}}} \right) \ll 1$. It should be stressed that this Abelian version of the Pagels-Tamboulis-like model has its validity limited to electric-type dominated configurations, for which 
\begin{equation}
{F_{\mu \nu }}{F^{\mu \nu }} =  - 2\left( {{{\bf E}^2} - {{\bf B}^2}} \right) < 0. \label{EHM10-a}
\end{equation}
We are actually interested in computing electrostatic interactions and the present
model with the condition ${F^2} < 0$ fits for our purposes.

It should, however, be noted here that this expression is analogous to that encountered in an Euler-Heisenberg-like electrodynamics at strong fields \cite{Nonlinear2}. We thus obtain a new equivalence between effective Abelian models. Notwithstanding, in order to put our discussion into context it is useful summarize the relevant aspects of the analysis described previously \cite{Nonlinear2}. Thus, our first undertaking is to introduce an auxiliary field, $\xi $, in order to handle the second term on the right hand in equation (\ref{EHM10}). This allows us to write the effective Lagrangian density as
\begin{equation}
{\cal L} = - \frac{1}{4}{\alpha _1}{F_{\mu \nu }}{F^{\mu \nu }} - {\alpha _2}{\left( {{F_{\mu \nu }}{F^{\mu \nu }}} \right)^2}, \label{EHM15}
\end{equation}
where ${\alpha _1} = 1 - 2\delta \left( {1 + \ln \xi } \right)$ and ${\alpha _2} = - \frac{{\delta \xi }}{{4{\Lambda ^4}}}$.

Analogously, to manipulate the quadratic term in equation (\ref{EHM15}) we introduce a second auxiliary field, $\eta$. In this manner, we then have 
\begin{equation}
{\cal L} =  - \frac{1}{4}\sigma {F_{\mu \nu }}{F^{\mu \nu }} + \frac{1}{{64{\alpha _2}}}{\left( {\sigma  - {\alpha _1}} \right)^2}, \label{EHM20}
\end{equation}
where $\sigma  = {\alpha _1} + 4{\alpha _2}\eta$.  
 
We are now in a position to calculate the expectation value of the energy operator $ H$ in the physical state $ |\Phi\rangle$. This calculation proceeds exactly as in the previous section. With this in view, the canonical momenta read ${\Pi ^\mu } =  - \sigma {F^{0\mu }}$, and at once we recognize the two primary constraints $\Pi ^0  = 0$ and ${\cal P}_\sigma \equiv \frac{{\partial L}}{{\partial \dot \sigma}} = 0$. Accordingly, the canonical Hamiltonian is expressed as
\begin{equation}
{H_C} = \int {{d^3}x} \left\{ {{\Pi ^i}{\partial _i}{A_0} + \frac{1}{{2\sigma }}{{\bf \Pi} ^2} + \frac{\sigma }{2}{{\bf B}^2} - \frac{1}{{64{\alpha}_2}}{{\left( {\sigma  - {\alpha}_1} \right)}^2}}\right\}. \label{EHM25}
\end{equation}
Time conservation of the primary constraint $\Pi^{0}$ yields the secondary constraint $\Gamma _1 \equiv \partial _i \Pi ^i  = 0$. Similarly for the ${\cal P}_\sigma$ constraint yields no further constraints and just determines the field $\sigma$. In this case, $\sigma$ is given by 
\begin{equation}
\sigma  = \left( {1 - 2\delta \left( {1 + \ln \xi } \right) + \frac{{4\delta {{\bf B}^2}}}{{{\Lambda ^4}}}\xi } \right)\left[ {1 + \frac{{12\delta {{\bf \Pi} ^2}}}{{{\Lambda ^4}}}\frac{\xi }{{{{\left( {1 - 2\delta \left( {1 + \ln \xi } \right) + \frac{{4\delta {{\bf B}^2}}}{{{\Lambda ^4}}}\xi } \right)}^3}}}} \right]. \label{EHM30}
\end{equation}

Hence we obtain
\begin{equation}
{H_C}=\int {{d^3}x} \left\{ {{\Pi _i}{\partial ^i}{A_0} + \frac{1}{2}{{\bf \Pi} ^2} + \delta \left( {1 + \log \xi } \right){{\bf \Pi} ^2} - \frac{{12\delta \xi }}{{{\Lambda ^4}}}{{\bf \Pi} ^4}} \right\}. \label{EHM35}
\end{equation}
Again, as in the previous section, we have ignored the magnetic field in equation (\ref{EHM35}) because it add nothing to the static potential calculation.
Next, requiring the primary constraint ${\cal P}_\xi$ to be preserved in time, one obtains the auxiliary field $\xi$. In this case $\xi  = {{{\Lambda ^4}} \mathord{\left/{\vphantom {{{\Lambda ^4}} {12{\Pi ^2}}}} \right.\kern-\nulldelimiterspace} {12{{\bf \Pi} ^2}}}$. Consequently, we get
\begin{equation}
 {H_C} = \int {{d^3}x} \left\{ {{\Pi _i}{\partial ^i}{A_0} + \frac{1}{2}\left( {1 + \frac{{16}}{3}\delta } \right){\Pi ^2} - \frac{{96\delta }}{{{\Lambda ^4}}}{\Pi ^4}} \right\}. \label{EHM40}
\end{equation}

As before, the corresponding total (first-class) Hamiltonian that generates the time evolution of the dynamical variables is $  
H = H_C  + \int {d^3 x} \left( {u_0(x) \Pi_0(x)  + u_1(x) \Gamma _1(x) } \right)$, where $u_o(x)$ and $u_1(x)$ are the Lagrange multiplier utilized to implement the constraints. Moreover, it follows from this Hamiltonian that $\dot A_0 \left( x \right) = \left[ {A_0 \left( x \right),H} \right] = u_0 \left( x \right)$, which is completely arbitrary function. Since ${\Pi ^0} = 0$ always, we discard both $A_0$ and $\Pi_0$ from the theory. Thus the Hamiltonian is now given as
\begin{equation}
H = \int {{d^3}x} \left\{ {w\left( x \right){\partial _i}{\Pi ^i} + \frac{1}{2}\left( {1 + \frac{{16}}{3}\delta } \right){\Pi ^2} - \frac{{96\delta }}{{{\Lambda ^4}}}{\Pi ^4}} \right\}, \label{EHM45}
\end{equation}
where $w(x) = u_1 (x) - A_0 (x)$.

We can evaluate the interaction energy by mimicking what we did previously. This then implies that 
the static potential for two opposite charges located at ${\bf y}\prime$ and $\bf y$ can be written as
\begin{equation}
V = \frac{{{e^2}}}{{4\pi }}\left( {1 - \frac{{16}}{3}\delta } \right)\log \left( {\frac{L}{{{{\tilde L_0}}}}} \right) + \frac{{12{e^4}}}{{{\pi ^3}}}\frac{\delta }{{{\Lambda ^4}}}\frac{1}{{{L^2}}},  \label{EHM50}
\end{equation}
with $|{\bf y} - {{\bf y}^ \prime }|=L$ and ${\tilde L_0}$ is a cut-off. It is straightforward to check that in the limit $\delta= 0$, expression (\ref{EHM50}) reduces to the Coulomb potential. However, for $\delta  < \frac{3}{{16}}$, the static profile for generalized Born-Infeld and Pagels-Tomboulis-like electrodynamics are equivalent.

Here, an interesting matter comes out. It is worth stressing that the previous static potential profile hinges crucially on the exponent in the second term on the right-hand side of equation (\ref{EHM15}). In other words, we call attention to the fact that the exponent in the term ${\left( {{F_{\mu \nu }}{F^{\mu \nu }}} \right)^2}$ be exactly $2$. In fact, in our previous work \cite{PatJose}, we have considered a phenomenologically effective model with exponent ${\raise0.5ex\hbox{$\scriptstyle 1$}\kern-0.1em/\kern-0.15em
\lower0.25ex\hbox{$\scriptstyle 2$}}$ and found a radically different result than the corresponding with exponent $2$. This requires a reconsideration of the connection between the specific value of the exponent and the corresponding potential, as we are going to illustrate in the next subsection.

\subsection{Related non-linear model}

As already stated, our next undertaking is to use our earlier procedure in order to examine the connection between the value of the exponent and the nature of the potential. The interest in this question emerges from the connection between scale symmetry breaking and confinement \cite{PatJose,GaeteGue1,GaeteGue2,GaeteGueSpa}, as well as from the one-loop effective action for QCD, where this new QCD vacuum acts like a dielectric medium and leads to confinement \cite{Dittrich}. 

Let us start off our considerations by considering the following three-dimensional Lagrangian density:
\begin{equation}
{\cal L} =  - \frac{1}{4}{F_{\mu \nu }}{F^{\mu \nu }} - M{\left( { - {F_{\mu \nu }}{F^{\mu \nu }}} \right)^p}, \label{ANLM05}
\end{equation}
where the $M$ constant has ${\left( {mass} \right)^{3\left( {1 - p} \right)}}$ dimension in natural units. As already expressed we confine ourselves to the domain $0 < p < 1$.

According to our procedure, we will introduce an auxiliary field, $v$, to handle the exponent in the Lagrangian (\ref{ANLM05}). Expressed in terms of this field, the Lagrangian (\ref{ANLM05}) takes the form
\begin{equation}
{\cal L} =  - \frac{1}{4}\left( {1 - \frac{{4Mp}}{{\left( {1 - p} \right)}}} \right){F_{\mu \nu }}{F^{\mu \nu }} - M{\left( {1 - p} \right)^{\frac{1}{{\left( {1 - p} \right)}}}}\frac{1}{{{v^{\frac{p}{{\left( {1 - p} \right)}}}}}}. \label{ANLM10}
\end{equation}
By introducing the auxiliary field $\frac{1}{V} \equiv 1 - \frac{{4Mp}}{{\left( {1 - p} \right)}}v$, expression  (\ref{ANLM10}) then becomes
\begin{equation}
{\cal L} =  - \frac{1}{4}\frac{1}{V}{F_{\mu \nu }}{F^{\mu \nu }} - {M^{\frac{1}{{\left( {1 - p} \right)}}}}\left( {1 - p} \right){\left( {4p} \right)^{\frac{p}{{\left( {1 - p} \right)}}}}{\left( {\frac{V}{{V - 1}}} \right)^{\frac{p}{{\left( {1 - p} \right)}}}}. \label{ANLM11}
\end{equation}

This new effective theory provide us with a suitable starting point to study the interaction energy. For this purpose, we start by observing that the canonical momenta read ${\Pi ^\mu } =  - \frac{1}{V}{F^{0\mu }}$, which produces two primary constraints $\Pi ^0=0$ and ${{\cal P}_v} = 0$. The canonical Hamiltonian is then
\begin{equation}
{H_C} = \int {{d^2}x} \left\{ {{\Pi ^i}{\partial _i}{A_0} - \frac{V}{2}{\Pi _i}{\Pi ^i} + \frac{1}{{4V}}{F_{ij}}{F^{ij}} + {M^{\frac{1}{{\left( {1 - p} \right)}}}}\left( {1 - p} \right){{\left( {4p} \right)}^{\frac{p}{{\left( {1 - p} \right)}}}}{{\left( {\frac{V}{{V - 1}}} \right)}^{\frac{p}{{\left( {1 - p} \right)}}}}} \right\}.    \label{ANLM15}
\end{equation}

Time conservation of the primary constraint $\Pi ^0$ leads to the secondary constraint $\Gamma _1 \equiv \partial _i \Pi ^i  = 0$. Likewise, for the constraint ${{\cal P}_v}$, we get the auxiliary field, $V$, satisfies the equation  
\begin{equation}
{V^{\left( {2p - 1} \right)}} - \frac{1}{{{2^{\left( {1 - p} \right)}}Mp}}{\left( {{{\bf \Pi} ^2}} \right)^{\left( {1 - p} \right)}}\left( {V - 1} \right) = 0.  \label{ANLM15}
\end{equation}
Evidently, to know the explicit form of V we have to choose $p$. In order to do so our considerations will be confined to the $p = {\raise0.5ex\hbox{$\scriptstyle 1$}\kern-0.1em/\kern-0.15em
\lower0.25ex\hbox{$\scriptstyle 2$}}$ and $p = {\raise0.5ex\hbox{$\scriptstyle 3$}
\kern-0.1em/\kern-0.15em\lower0.25ex\hbox{$\scriptstyle 4$}}$ cases.

Following the same steps as in the previous section, the extended Hamiltonian that generates translations in time then reads
\begin{equation}
H = \int {{d^2}x} \left\{ {w\left( x \right){\partial _i}{\Pi ^i} + \frac{V}{2}{{\bf \Pi} ^2} + {M^{\frac{1}{{\left( {1 - p} \right)}}}}\left( {1 - p} \right){{\left( {4p} \right)}^{\frac{p}{{\left( {1 - p} \right)}}}}{{\left( {\frac{V}{{V - 1}}} \right)}^{\frac{p}{{\left( {1 - p} \right)}}}}} \right\},    \label{ANLM20}
\end{equation}
where $w(x) = u_1 (x) - A_0 (x)$ and the auxiliary field $V$ satisfies equation (\ref{ANLM15}).

Next, since our main motivation is compute the static potential for the model under consideration, we shall adopt the same gauge-fixing condition that was used in the previous section. Therefore, the fundamental Dirac bracket is given by expression (\ref{EHM45}).

We now have all the information required to compute the potential energy for this theory. To do this, we will use the gauge-invariant scalar potential which is given by expression (\ref{BIL100}).

It follows from the above discussion that Gauss' law takes the form
\begin{equation}
\nabla  \cdot {\bf \Pi} = 0, \label{ANLM25}
\end{equation}
where ${\bf \Pi}$ is given by
\begin{equation}
{\bf \Pi}  = \frac{1}{V}{\bf E}. \label{ANLM30}
\end{equation}
Again, we see that, for ${J^0}\left( {t,{\bf r}} \right) = e{\delta ^{\left( 2 \right)}}\left( {\bf r} \right)$, the ${\bf \Pi}$-field lies along the radial direction and is given by ${\bf \Pi} = \frac{e}{{2\pi {r}}}\hat r$. Making use of this result, we find that the electric field, for $p = {\raise0.5ex\hbox{$\scriptstyle 1$}
\kern-0.1em/\kern-0.15em\lower0.25ex\hbox{$\scriptstyle 2$}}$, reduces to
\begin{equation}
{\bf E} = \frac{Q}{r}\left( {1 + \frac{{\sqrt 2 M}}{Q}r} \right)\hat r. \label{ANLM35}
\end{equation}
Whereas, for $p = {\raise0.5ex\hbox{$\scriptstyle 3$}\kern-0.1em/\kern-0.15em\lower0.25ex\hbox{$\scriptstyle 4$}}$, the electric field becomes
\begin{equation}
{\bf E} = \frac{Q}{r}\left( {1 + \frac{9}{{2\sqrt 2 }}\frac{{{M^2}}}{Q}r \pm \frac{9}{{2\sqrt 2 }}\frac{{{M^2}}}{Q}r\sqrt {1 + \frac{4}{{9\sqrt 2 }}\frac{Q}{{{M^2}}}\frac{1}{r}} } \right)\hat r.  \label{ANLM40}
\end{equation}

Finally, a procedure similar to that used in the preceding section allows us to obtain the static potential for two opposite charges located at $\bf 0$ and $\bf L$. For $p = {\raise0.5ex\hbox{$\scriptstyle 1$}
\kern-0.1em/\kern-0.15em\lower0.25ex\hbox{$\scriptstyle 2$}}$, the static potential is then 
\begin{equation}
V = \frac{{{e^2}}}{{4\pi }}\ln \left( {\frac{L}{{{{\bar L}_0}}}} \right) + \frac{{eM}}{{\sqrt 2}} L,  \label{ANLM45} 
\end{equation}
where ${\bar L}_0$ is a cut-off. The above potential profile is analogous to the one encountered, by a different method, for Yang-Mills and a Born-Infeld term \cite{PatJose}.

Whereas, for $p = {\raise0.5ex\hbox{$\scriptstyle 3$}\kern-0.1em/\kern-0.15em\lower0.25ex\hbox{$\scriptstyle 4$}}$, the static potential profile becomes
\begin{equation}
V = \frac{{{e^2}}}{{4\pi }}\ln \left( {\frac{L}{{{{\bar L}_0}}}} \right) + \frac{{9e{M^2}}}{{4\sqrt 2 }}L \pm \left[ {\frac{{M{e^{{\raise0.5ex\hbox{$\scriptstyle 3$}
\kern-0.1em/\kern-0.15em
\lower0.25ex\hbox{$\scriptstyle 2$}}}}}}{{3\sqrt \pi  {2^{{\raise0.5ex\hbox{$\scriptstyle 9$}
\kern-0.1em/\kern-0.15em
\lower0.25ex\hbox{$\scriptstyle 4$}}}}}}\sqrt L \sqrt {1 + \frac{{9{M^2}\pi }}{{{2^{{\raise0.5ex\hbox{$\scriptstyle 1$}
\kern-0.1em/\kern-0.15em
\lower0.25ex\hbox{$\scriptstyle 2$}}}}e}}L}  + \frac{{{e^2}}}{{4\pi }}Arc{\mathop{\rm Sinh}\nolimits} \left( {\frac{{3M\sqrt \pi  }}{{{e^{{\raise0.5ex\hbox{$\scriptstyle 1$}
\kern-0.1em/\kern-0.15em
\lower0.25ex\hbox{$\scriptstyle 2$}}}}{2^{{\raise0.5ex\hbox{$\scriptstyle 1$}
\kern-0.1em/\kern-0.15em
\lower0.25ex\hbox{$\scriptstyle 4$}}}}}}\sqrt L } \right)} \right].  \label{ANLM50}
\end{equation}
The above result clearly illustrates that the presence of the ${\left( { - {F_{\mu \nu }}{F^{\mu \nu }}} \right)^{{\raise0.5ex\hbox{$\scriptstyle 3$}\kern-0.1em/\kern-0.15em\lower0.25ex\hbox{$\scriptstyle 4$}}}}$ term also leads to confinement at Abelian level, when compared to the one-loop effective action for QCD in $(2+1)$-dimensions \cite{Dittrich}.

\section{Final Remarks}

Let us summarize our work. Using the gauge-invariant but path-dependent variables formalism, we have
computed the static potential for three different three-dimensional field theoretic models. Once again, a correct identification of physical degrees of freedom has been fundamental for understanding the physics hidden in gauge theories. It was shown that generalized Born-Infeld and Pagels-Tomboulis electrodynamics are equivalent. In this way we have provided a new connection between effective models. Interestingly enough, for an Euler-Heisenberg-like electrodynamics the interaction energy contains a linear potential, leading to the confinement of static charges. However, it is interesting to emphasize that this linear potential is obtained in an Abelian model in. Finally, the benefit of considering the present framework is to provide unifications among different models.

\section{ACKNOWLEDGMENTS}
P. G. was partially supported by Fondecyt (Chile) grant 1130426 and by Proyecto Basal FB 0821. P. G. also wishes to thank the Field Theory Group of the CBPF for hospitality.

\end{document}